\documentclass[twocolumn]{revtex4}
\usepackage{epsfig}
\usepackage{graphicx}

\begin{document}
\title{Magnetic polarons and magnetoresistance in EuB$_6$}
\author{M.J. Calder\'on$^{1}$, L.G.L. Wegener$^{1}$, and P.B. Littlewood$^{1,2}$}
\affiliation{
(1) Cavendish Laboratory, Cambridge University, Madingley Road,
Cambridge CB3 0HE, UK\\
(2) National High Magnetic Field Laboratory, Pulsed Field Facility, LANL, Los Alamos NM 87545, USA}
\date{\today}

\begin{abstract}
EuB$_6$ is a low carrier density ferromagnet which exhibits large
magnetoresistance, positive or negative depending on temperature.
The formation of magnetic polarons just above the magnetic
critical temperature has been suggested by spin-flip Raman scattering experiments.
We find that the fact that EuB$_6$ is a semimetal has to be taken into account
to explain its electronic properties, including magnetic polarons and 
magnetoresistance. 
\end{abstract}

\maketitle

\section{Introduction}
Europium hexaboride is part of the large and heterogeneous class of
materials that exhibit Colossal Magnetoresistance (CMR). The ferromagnetic
transition in EuB$_6$ is accompanied by a dramatic change in resistivity.
There is a large body of experimental data available on the magnetic and
electric properties, but a thorough understanding 
is lacking.

EuB$_6$ has a cubic unit cell with Eu-ions at its vertices and a Boron
octahedron at its center.  The material is ferromagnetic and shows two
magnetic transitions: at $T_\mathrm{M} = 15.3$K and at $T_\mathrm{C}=
12.5$K \cite{sullow98}. These have been associated with a spin
reorientation and a ferromagnetic transition, respectively. 
Neutron diffraction experiments \cite{tarascon81}
have given a magnetic moment $\mu= 7.3 +0.5 \mu_B$.
This is exclusively due to the localized half-filled $f$-shell in
the Eu$^{+2}$ ions.\cite{massida97}

Electronic structure calculations \cite{massida97}, Shubnikov-de Haas
and de Haas-van Alphen measurements \cite{aronson99} show that EuB$_6$
is a semimetal. The Fermi surface consist of two ellipsoidal pockets,
one electron-like and one hole-like,
centered on the $X$ point in the Brillouin zone. The pockets contain very
few carriers: Hall effect measurements yield $n_\mathrm{eff} \sim 10^{-2}$
carriers per formula unit \cite{paschen00} at low temperatures. 
Small dilations of the boron octahedra cause overlap of the conduction and valence bands at the $X$ points rendering EuB$_6$ semimetallic.
The carrier
concentration decreases smoothly as temperature increases.

The electrical resistivity is metallic in the ferromagnetic regime. It
shows a sharp peak near $T_\mathrm{C}$. Above this temperature, the
resistivity decreases with an almost activated temperature dependence
until it reaches a minimum at about 30K. At higher temperatures it
increases and eventually starts to saturate at about room temperature.

The application of a magnetic field produces sharp changes in the resistivity.
Close to the magnetic transition, 
{\em negative} magnetoresistance (MR) values of up to
100\% have been observed \cite{paschen00,sullow00}. This decrease in
resistivity is accompanied by a large decrease in the (negative) 
Hall coefficient\cite{paschen00}
and an increase in the plasma frequency \cite{broderick02}. The change in the
plasma frequency is more gradual than the changes in resistivity and
Hall coefficient.

In the ferromagnetic regime, on the other hand, the MR is large and {\em positive}:
at 1.7 K resistivity changes of up to 700\% have been observed in
a transversal applied field of 7T \cite{paschen00,aronson99}. The MR depends
quadratically on the applied field strength at low temperatures\cite{wigger03}.

Just above T$_C$ and up to $\sim 30K$, 
the existence of magnetic polarons has been proposed as
the cause of the Stokes' shift measured with
Spin Flip Raman Scattering (SFRS)\cite{nyhus97,snow01}. The resistivity
is activated at these temperatures. However, this data contains a previously unremarked
puzzle, in that
the energy scale turns out to be considerably lower - by a factor of thirty - than expected based on reliable
estimates of the exchange interaction.

Wigger et al.\cite{wigger03} showed how the crossover between large positive and 
large negative MR from well below to well above the ferromagnetic transition can be explained
by the dominance of orbital scattering at $T \ll T_c$ to spin scattering at $T \gg Tc$. The model
we shall use for the carrier transport in these regimes is similar to that of ref.\cite{wigger03} and
we shall thus suppress most of the details. The key feature of the model is its multiband nature - there are two
types of carrier.

In this paper we concentrate principally on the regime close to $T_c$ and
analyze the evidence for the existence of magnetic polarons in Europium Hexaboride. 
We show how the SFRS results can be explained using a multiband model, resolving the conundrum
of the anomalously small energy associated with the carrier spin flip.

\section{Model and parameters}
We model EuB$_6$ as a ferromagnetic semimetal with a low carrier density.
Both electrons and holes are itinerant and are coupled to the local moments 
$S=7/2$. This can be described by the following general 
Hamiltonian:

\begin{eqnarray}
H &=& -t \sum_{i,j,\sigma} c_{i\sigma}^+ c_{j\sigma}
   -J \sum_{i,j} {\vec{S}_i} {\vec{S}_j} \nonumber\\
    &-& J'_e\sum_i  c_{i\sigma e}^+ {\vec{\tau}}_{\sigma,\sigma'} c_{i\sigma e}{\vec{S}_i}
    -J'_h\sum_i  c_{i\sigma h}^+ {\vec{\tau}}_{\sigma,\sigma'} c_{i\sigma h}{\vec{S}_i}
\label{eq:hamiltonian}
\end{eqnarray}

Here, the hopping parameter is roughly $t=0.1$ eV\cite{massida97}.
$c_{i\sigma (e,h)}^+ {\bf \tau}_{\sigma,\sigma'} c_{i\sigma (e,h)}$ is the itinerant
carrier spin operator and the subindices $e$ and $h$ stand for electrons 
and holes respectively.
$J'_e$ ($J'_h$) is the on-site coupling between the spins of the electrons (holes) and
the local moments. 
$J$ is the magnetic exchange between local moments.

First of all we need to discuss what is the origin of the ferromagnetism 
and the order of magnitude of the magnetic couplings. 
Ferro- and antiferro-magnetism of the insulating Eu-chalcogenides (EuX, X$=$ O, S, Se, Te) has been explained
as due to superexchange interaction between neighbor Eu ions 
\cite{methfessel,goodenoughBook} through the anion between them. The density
of carriers in the undoped chalcogenides is too low to expect any indirect RKKY
(Ruderman-Kittel-Kasuya-Yosida) 
interaction. The ferromagnetic interaction arises instead from the overlap 
between the 4$f$- and 5$d$-orbitals at different cations. This overlap 
leads to an effective exchange interaction in third order in 
perturbation theory\cite{goodenoughBook}. This does not  
apply directly to EuB$_6$
due to the different crystalline structure, but nevertheless one 
expects that the superexchange coupling $J$ is small. Moreover, the increase of 
 magnetic critical temperature and concomitant decrease of resistivity under high 
pressures \cite{cooley97} has revealed    
that the magnetic exchange in EuB$_6$ is mainly due to the RKKY
interaction. Therefore, 
$J$ in Eq.~\ref{eq:hamiltonian} is negligible.

The RKKY magnetic exchange is mediated by 
the itinerant carriers via their coupling with the lattice
magnetic moments. An effective Heisenberg-like magnetic exchange can
be written in terms of the local Hund's like exchange coupling $J'$ 
($J'_e$ or $J'_h$) \cite{rkky}
\begin{equation}
J_{eff}=-9 \pi  {\frac{J'^2}{E_F}} n^2 \sum_i F(2 k_F r_i)
\label{eq:RKKY}
\end{equation}
where $F(x)={\frac{-x \cos x +\sin x}{x^4}}$, $n$ is the density
of carriers, and $E_F$ is the Fermi energy. $J_{eff}$ is an
oscillating function of $x$ but is ferromagnetic for small $x$.
This is the relevant limit for Europium Hexaboride, as its
low carrier density implies $k_F r \rightarrow 0$.

To estimate the value of $J'$ we use the mean field relation between
T$_C$ and J$_{eff}$,  $T_c \sim z S^2 J_{eff}$,
where  $z$ is the coordination number for Eu, and $S$ is the
$z$-component of the local moments. Using a critical temperature
$T_c \sim 12 \,\, K$ and a parabolic approximation to the bands, 
$ J' \sim 0.1\,\, eV$ is found, consistent
with reported data for isolated Eu-ions\cite{russel41}. In this estimation
we are considering that only one kind of carrier is responsible for the magnetism.

\section{Magnetic polarons}

When the local exchange coupling $J'$ is large enough, carriers  
can be localized by ferromagnetic clusters and form composite 
objects called magnetic polarons. Ferromagnetic polarons
can exist in the low temperature phases of antiferromagnets
but here we are interested in those formed in the paramagnetic
phase.
A necessary condition for the existence of magnetic polarons
is that the density of carriers is very low compared to the 
inverse of the correlation volume, namely $n\xi^3 <<1$. When
this condition is fulfilled, polarons are well-defined 
non-overlapping entities.

There are two kinds of magnetic polarons: free and bound. A free
magnetic polaron is a carrier localized in a ferromagnetic cluster 
embedded in a paramagnetic background. 
A carrier that is coupled strongly to local moments
via a Hund's like coupling tends to align the moments that are
within a Bohr radius. This causes a trapping potential that localizes
the carrier. The potential can be enhanced by random fluctuations
of the magnetization that produce an alignment of local moments in
the carrier's vicinity\cite{dietl,heiman83}.

The carrier thus traps itself by the magnetization it causes. It
could increase the alignment of the local moments and hence decrease
its energy by localizing itself in a smaller volume. However, this
would lead to an increase in kinetic energy.  The quantity that
determines the stability and size of these objects is therefore
$J'/t$ where $J'$ is the coupling of the carrier spin to the local moments
and $t$ is the hopping parameter.  The ratio $J'/t$ needs
to be typically larger than one \cite{majumdarPRL98,calderon00} to guarantee
stability of the free magnetic polaron. 

On the other hand, in bound magnetic
polarons the main driving force
trapping the carrier is not the local magnetic interaction but the
electrostatic potential created by impurities.  The formation of
the ferromagnetic cluster described above does occur. However, it
is a second order process, as the magnitude of Hund's coupling is
much smaller than the Coulomb interaction.

Mean field \cite{majumdarPRL98} and Monte-Carlo \cite{calderon00} calculations
have shown that magnetic polarons can exist within a temperature window
above T$_C$ whose width depends on the ratio $J'/t$. At higher temperatures,
magnetic fluctuations are strong enough to destroy the magnetic polarons.
Below $T_C$, the condition $n\xi^3<<1$ is not fulfilled and the 
polarons overlap. 
If a magnetic field is
applied within the existence temperature window, 
the size of a polaron increases until eventually
the polarons overlap and produce a ferromagnetic transition. 

Free and bound magnetic polarons can be differentiated by their dynamics
and the resistivity they cause. Bound magnetic polarons are bound
to an impurity in the system so the only way of transport is via an 
activated process: when the trapped carrier is ``ionized'' it is free
to move until it is trapped by another impurity. Therefore they produce
a resistivity $\rho$ such that $\partial \rho/ \partial T <0$. In contrast, 
free magnetic
polarons are able to move to adjacent areas when random fluctuations of the 
nearby spins produce an aligned region. There is not a barrier to overcome
in this process. This transport mechanism has been called 
``fluctuation-induced 
hopping'' \cite{wegener02} and produces a metallic resistivity  
$\partial \rho/ \partial T >0$.  

Magnetic polarons have been largely studied in connection with 
Eu-chalcogenides (EuO, EuS, EuSe, EuTe) \cite{kasuya68,kasuya70,vonmolnar67} 
and diluted magnetic semiconductors
such as Cd$_{1-x}$Mn$_x$Te and Pb$_{1-x}$Mn$_x$Te with $x$, the concentration 
of magnetic ions, small. Experimental evidence included
photoluminescence spectra \cite{golnik83} and magneto-optical experiments
as Spin Flip Raman Scattering \cite{nawrocki81,heiman83,isaacs88}. 
The Raman scattering spectrum shows an inelastic peak at low frequencies
(Stokes' shift) which, for the diluted magnetic semiconductors, depends as 
follows on polaronic properties:\cite{dietl}
\begin{equation}
\Delta E= \bar x J' \langle S_z \rangle,
\label{eqn:stokesshift}
\end{equation}
where $\bar x$ is the density of magnetic ions participating
in the formation of polarons \cite{note1}, and $J'$ is the local
exchange interaction between the $s$ itinerant electrons and the
$d$ electrons localized in the Mn ions. The low density of local moments makes 
for a small Stokes' shift, which - for Cd$_{1-x}$Mn$_x$Se - is consistent 
with experiment\cite{dietl}.

SFRS measurements done in EuB$_6$ have similarly revealed a zero-field peak 
in scattered 
intensity of the order of $\sim 12 meV$ \cite{nyhus97,snow01} at $18K$, just 
above the magnetic critical temperature. The behavior of this peak with 
temperature and external magnetic field is consistent with the stability
conditions theoretically established for magnetic polarons.
Free magnetic polarons are not expected to be stable in  EuB$_6$ as $J'$ 
and $t$ are comparable. Moreover, 
we have argued above that the activated behavior of the
resistivity is better explained by means of bound magnetic polarons.
Eu-site vacancies  
would produce the binding Coulomb potential for electrons.
    
We expect the energy of the Stokes' shift to be given by 
Eq.~\ref{eqn:stokesshift}
but now $\bar x\sim 1$ as EuB$_6$ has a local moment on every Eu site 
in the cubic 
lattice. Very few site vacancies are expected in this fairly clean material. 
Mean-field theory\cite{kasuya70}
predicts that bound magnetic polarons are fully spin polarised so
$\langle S_z \rangle=7/2$.  Using this value and the energy of the
Stokes' shift we obtain $ J' \sim 3$meV. This is far too low
compared to the values reported in the literature for $ J'$ in
isolated Eu ions $\sim 100$meV \cite{russel41,nyhus97}.

We are therefore left with a conundrum: the peak in the light scattering intensity
follows all the trends calculated for an object with magnetic origin but the energy 
of that peak is almost two orders of magnitude smaller
than expected. The solution to this problem lies in the fact that both polarised
electrons and holes are found at the Fermi Energy. 

Electrons and holes come from different
B and Eu orbitals and therefore their magnetic couplings to the localized
spin in the Eu $4f$ orbitals, $J'_e$ and $J'_h$ respectively, can be 
very different. Electronic structure calculations \cite{massida97} 
reveal that the hole pocket comes from the highest intraoctahedron B $2p$ band.
On the other hand, the electron pocket comes from bonding combinations of the cation 
$d$ orbitals pointing along the cartesian directions with some hybridization 
with the B atoms and some free-electron-like character on the (110) 
axes between the cations. In other words, the electron charge density distribution
is mainly found around the Eu ions while the holes are found around the B.  
Therefore, the coupling of the electrons is 
expected to be much larger than that of the holes. Consistently, 
Fig. 10 in Ref.\cite{massida97}
shows a much larger majority-minority spin band splitting for electrons than for holes.

In conclusion, we propose that the ferromagnetic ordering is produced by the 
itinerant electrons coupled to the localized spin in Eu with $J'_e\sim 100 meV$, while the itinerant holes,
much more weakly coupled ($J'_h\sim 5meV$), account for the SFRS Stokes' shift.
A corollary of this identification is that there is likely a much higher energy feature
in the SFRS, so far unobserved, that correspond to spin-flip of the electron state.
 
In the following section we will see how the existence of both electrons and holes
is necessary to explain other electronic properties of EuB$_6$.

\section{Magnetoresistance}

\subsection{Positive MR at $T < T_c$}

At low temperatures the magnetoresistance is due to the presence of two
types of carriers and we will call it ``orbital'' MR. There are two 
effects involved. The same physics that causes the Hall effect is the most 
important cause of the MR. In addition to this, there is also a small shift 
of the bands with applied magnetic field that causes a small change in the 
carrier density.

In a simple metal with one type of carrier and a simple Fermi
surface, there is no MR.  A current that flows perpendicular to a
magnetic field is initially deflected due to the transverse Lorentz
force.  This causes an electric field, the Hall field $\vec{E}_\mathrm{H}$,
to build up:
\begin{equation}
\vec{E}_\mathrm{H} = \frac{e \tau}{4 \pi m} \vec{H}\times\vec{J}.
\label{eqn:Hall_Voltage}
\end{equation}
The carriers that are deflected hit the edge of the sample
and accumulate. The field that is thus built up counteracts the Lorentz
force.  When it  cancels the Lorentz force, the current is undeflected.
This is well known; it means that there is no MR in a normal metal.
The component of the current density that is parallel to the applied
electric field is not affected by the magnetic field. Therefore
the resistivity remains unaffected as well.

In a semimetal, on the other hand, there are by definition two
kinds of carriers. These two kinds will almost invariably have
different scattering times and different masses. Eq.~\ref{eqn:Hall_Voltage} 
shows that the Hall voltage depends on the
scattering time and the mass of the carrier.  Therefore, the Hall
voltages of the different kinds of carriers are also different.
There is thus no voltage at which the two carriers will travel
through the sample without deflection. 

The difference in the Hall voltages is proportional to the applied
magnetic field, so that the component of the current parallel to
the applied electric field decreases with increasing field.  The
resistivity increases therefore when a magnetic field is applied,
and the MR is therefore positive. We call this orbital MR, since
it is due to the difference of the masses and scattering times of
the two types of carriers. These properties derive from particularities
of the atomic orbitals in the material.

The MR can be calculated from a linearized Boltzmann equation that includes 
a magnetic field under the assumption of a spherical Fermi surface. 
This is a standard calculation, which can be found in  
\cite{ziman} and \cite{abrikosov}.
We quote the result for the magnetoresistance:
\begin{equation}
\frac{\Delta\rho(H)}{\rho_0} = 
\frac{\sigma_\mathrm{e}\sigma_\mathrm{h}(\mu_\mathrm{e}-\mu_\mathrm{h})^2 
    H^2}{(\sigma_\mathrm{e}+\sigma_\mathrm{h})^2 + H^2 
	(\mu_\mathrm{e}\sigma_\mathrm{e} + \mu_\mathrm{h}\sigma_\mathrm{h})^2},
\label{eqn:zimanMR}
\end{equation}
where $\sigma_\mathrm{e}$ and $\sigma_\mathrm{h}$ are the electron
and hole conductivity respectively, and $\mu_\mathrm{e}$ and
$\mu_\mathrm{h}$ are their mobilities, both in the absence of any
magnetic field. The right hand side of the above formula is easily seen to
vanish if both types of carriers have the same mass and scattering
time: the difference of the mobilities in the numerator vanishes
in that case.

At low temperatures the orbital MR is largest: the scattering time is
largest, so that the deflection is largest as well. As the scattering
time decreases the effect becomes less important. For sufficiently
small scattering times orbital MR becomes negligible.

% 2.2.2 Band shifting. I will refer back to this section when discussing
% critical MR
The second effect that causes MR in EuB$_6$ is the shifting of the bands 
when a magnetic field is applied. The shift is caused by the coupling $J'$ of 
the carriers to the Eu local moments. Let us write the local moments as the 
sum of their average and the deviations therefrom: 
$\vec{S}_i = \langle \vec{S}\rangle + \vec{\delta S}_i$. We now use this 
expression in the Hamiltonian in Eq.~\ref{eq:hamiltonian}, and obtain a 
term that couples the carrier's energy to the magnetization. As the 
magnetization grows, the electron-like band is shifted to lower 
energies, and the hole-like band to higher energies. This causes an increase 
in the number of carriers as carriers spill over from one band into the 
other. 

%The Fermi level shifts slightly as the external field is applied 
%since both bands have slightly different masses.  

The change in carrier density can be obtained from the following two 
requirements. Firstly, the band-shift introduced
by the change in magnetization $M$ is $\delta E = J'_eM/2$ for electrons
and $\delta E = J'_hM/2$ for holes. Secondly, the 
increase in the number of carriers of both kinds is equal: as the hole like 
band is shifted up, new holes are created as negatively charged particles spill 
into the electron-like band and vice versa. Overall charge balance is maintained. 
We also assume a spherical Fermi surface, an accurate assumption in the case of EuB$_6$ \cite{sullow98}.
We can use the requirement that charge neutrality be 
conserved to calculate the shift in the Fermi level. The shift is then used 
to obtain the number of carriers by integrating the density of states up to 
the Fermi level. The increase resulting from the shift of the bands is 
small, even at full saturation of the magnetization. Its effect on the MR 
at low temperatures is then negligible. 
The magnetization as a function of applied field is obtained from a 
Curie-Weiss model.

% 2.2.3 combine two effects
We included this change in carrier density due to the bands shifting
in our model for the  orbital MR. The contribution to the MR due
to band shifting is opposite to that of the orbital effects. An
external magnetic field increases the carrier density, so that
it decreases the resistivity.  Since the magnetization is almost
saturated at low temperatures, the carrier density does not change
much with applied field, and the MR is affected only slightly. The
orbital contribution to the MR dominates.

\begin{figure}
\scalebox{.4}{\includegraphics{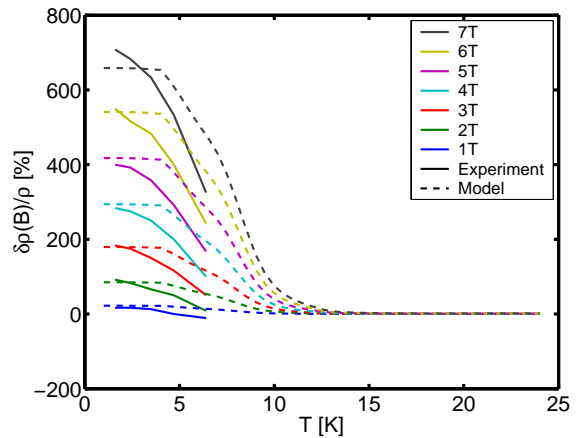}}
\caption{Orbital MR at low temperatures for different applied magnetic fields. The data is taken from
ref\cite{paschen00}. The plateau seen in the model below 5 K is an artifact of data digitization.  \label{fig:MRLowT}}
\end{figure}

We plot the MR obtained for the combined effect of the band shifts
and the orbital effects in Fig.~\ref{fig:MRLowT}. The temperature
dependent scattering times at zero field are obtained from experimental
data in \cite{paschen00}. The electron and hole mobilities are
obtained from the conductivity at zero field, the masses of the
carriers\cite{aronson99} and the carrier densities.  The latter
were obtained from the plasma frequency in \cite{paschen00}. We
introduced a small imbalance between the carrier densities of 
$6.10^{-4}q_\mathrm{e}$ per unit cell, in accordance with \cite{aronson99}.  
This imbalance is thought to arise from impurities. These numbers provide
input to the model.

Our simple model, which depends only on parameters
measured at zero field,  can reproduce the large positive MR at low temperatures 
accurately. Fig.~\ref{fig:MRLowT} shows that it reproduces the magnitude
of the MR well. It also predicts the (nearly) quadratic dependence in applied 
field. From Fig.~\ref{fig:MRLowT} it is clear that the orbital contribution to the 
MR dwindles at higher temperatures. 
The model proposed will lose its validity near the magnetic transition 
and in the paramagnetic phase, when other effects dominate.

\subsection{Negative MR at $T \approx T_c$ and above}

The mechanisms that govern the low-temperature MR become insignificant 
near the ferromagnetic transition. Close to $T_c$ the 
scattering time is so short that the positive orbital contribution to the 
MR is negligible. On the other hand, 
the shift of the bands caused by an applied magnetic field becomes 
substantial. We estimate that the carrier density changes by about $7\%$
as the applied field saturates the magnetization.
This could explain only part of the
negative MR in the critical regime.

Aditionally, near the critical point, spin fluctuations may provide
a large contribution to the electrical resistance.
The dominant modes near a ferromagnetic transition 
are those with small $q$. They only produce substantial backscattering
if $2k_F$ is itself small \cite{majumdar98,majumdarPRL98}. 
This is the case of EuB$_6$ as its carrier density is very small.
The suppression of the spin fluctuations when a magnetic field 
is applied is largely responsible for the MR found in the critical regime.
Moreover, as shown in Ref. \cite{majumdarPRL98} (see Fig. 3), the localization
of the carriers in magnetic polarons further increases the magnetoresistance. 

In the temperature regime - just above $T_c$ - where the SFRS data gives evidence for
bound magnetic polarons, the magnetoresistivity is large, and strongly negative, as expected since an applied magnetic field
suppressed the magnetic polarons. At temperatures greater than about $30K$ where the polarons are
destabilized, we have a smaller (but still negative) MR dominated by local spin fluctuation scattering.

As mentioned in the introduction, the negative MR is accompanied by a shift
in the plasma frequency \cite{broderick02}. 
The carrier density change produced by the band shifting alone
is too small to 
account for that shift.  This is in contrast to the results by 
Wigger {\it et al} \cite{wigger03} who claim consistency between the experiment
and their calculation, though the functional dependence 
$M^2 \propto (\omega_p^2)^2$ \cite{broderick02} is not reproduced. 
There are two main differences between their model and ours: they consider 
EuB$_6$ to be a strongly compensated n-type magnetic semiconductor and use
the same local coupling $J'$ for electrons and holes.

\section{Summary}

EuB$_6$ is a low carrier density ferromagnet with unusual properties:
the resistance changes from metallic to activated and then to metallic as
the temperature increases and the magnetoresistance changes sign close to 
$T_C$. The activated region has been ascribed to the existence of bound
magnetic polarons. We discuss their existence in the light of 
Spin Flip Raman Scattering measurements reported in Ref.\cite{nyhus97,snow01}. 
We conclude that the signature seen by those experiments is due to 
the Hund's like coupling of holes with the local spins while electrons
are responsible for the magnetic ordering through the RKKY interaction. This resolves
the puzzle that the RKKY transition temperature implies an exchange coupling of the carriers
to the local moment of about 0.1 eV, about 30 times larger than the measured spin-flip energy of
a carrier trapped in a bound polaron. The existence of bound magnetic polarons is also consistent with
a large {\em negative} magnetoresistance above $T_c$.
The {\em positive} magnetoresistance in the ferromagnetic phase is also produced by 
the interplay of two kinds of carriers with different masses and
scattering times.

MJC acknowledges Churchill College, University of Cambridge,  for financial support. 
The NHMFL is supported by the National Science Foundation, 
the state of Florida and the US Department of Energy.


\begin{thebibliography}{99}
\bibitem{note1} In the diluted magnetic semiconductors of the type 
Cd$_{1-x}$Mn$_x$Se,
the Mn are antiferromagnetically coupled via superexchange. This is a 
shortrange interaction.
At very low $x$, the probability of having Mn clusters is very low so 
$\bar x \simeq x$.
For larger $x$, antiferromagnetically coupled clusters of Mn form, 
so not all the Mn ions
will participate in the magnetization of the system. See, for instance 
\cite{heiman83}.
\end{thebibliography}
\end{document}